# Topological Band Engineering of Graphene Nanoribbons

**Daniel J. Rizzo[1,†], Gregory Veber[2,†], Ting Cao[1,3,†], Christopher Bronner[1], Ting Chen[1], Fangzhou Zhao[1], Henry Rodriguez[1], Steven G. Louie[1,3,*], Michael F. Crommie[1,3,4,*], Felix R. Fischer[2,3,4,*]**

[1]Department of Physics, University of California, Berkeley, CA 94720, USA. [2]Department of Chemistry, University of California, Berkeley, CA 94720, USA. [3]Materials Sciences Division, Lawrence Berkeley National Laboratory, Berkeley, CA 94720, USA. [4]Kavli Energy NanoSciences Institute at the University of California Berkeley and the Lawrence Berkeley National Laboratory, Berkeley, California 94720, USA.

[†] These authors contributed equally to this work.

* Corresponding authors

Topological insulators (TIs) are an emerging class of materials that host highly robust in-gap surface/interface states while maintaining an insulating bulk[1,2]. While most notable scientific advancements in this field have been focused on TIs and related topological crystalline insulators[3] in 2D[4-6] and 3D[7-10], more recent theoretical work has predicted the existence of 1D symmetry-protected topological phases in graphene nanoribbons (GNRs)[11]. The topological phase of these laterally-confined, semiconducting strips of graphene is determined by their width, edge shape, and the terminating crystallographic unit cell, and is characterized by a $Z_2$ invariant (similar to quasi-1D solitonic systems[12-15]). Interfaces between topologically distinct GNRs characterized by different $Z_2$ are predicted to support half-filled in-gap localized electronic states which can, in principle, be utilized as a tool for material engineering[11]. Here we present the rational design and experimental realization of a topologically-engineered GNR superlattice that hosts a 1D array of such states, thus generating otherwise inaccessible electronic structure. This strategy also enables new end states to be engineered directly into the termini of the 1D GNR superlattice. Atomically-precise topological GNR superlattices were synthesized from molecular precursors on a Au(111) surface under ultra-high vacuum (UHV) conditions and characterized by low temperature scanning tunneling microscopy (STM) and spectroscopy (STS). Our experimental results and first-principles calculations reveal that the frontier band structure of these GNR superlattices is defined purely by the coupling between adjacent topological interface states. This novel manifestation of 1D topological phases presents an entirely new route to band engineering in 1D materials based on precise control of their electronic topology, and is a promising platform for future studies of 1D quantum spin physics.

**Keywords:** symmetry protected topological materials, topological insulators (TIs), graphene nanoribbons (GNRs), band engineering, organic semiconductors

GNRs represent a privileged scaffold in the exploration of novel topological phases because graphene becomes semiconducting when confined laterally with certain edge structures[16,17]. Recent advancements in the rational bottom-up synthesis of GNRs have provided atomically precise control over virtually all structural parameters through the rational design and self-assembly of small-molecule precursors[18]. This has allowed exploration of GNR energy-gap vs. width relations[16,17] and bandgap engineering via dopant-mediated shifts in electron affinity[19-24]. Fusion of different types of GNR precursors along the longitudinal axis has led to the design and synthesis of type I and type II heterojunctions where GNR electronic structure changes continuously from one GNR type to another as the heterojunction interface is crossed[25,26].

Topological concepts, on the other hand, provide an entirely new strategy in the design of bottom-up GNR electronic structure. We herein exploit the nontrivial topological phases determined by a $Z_2$ invariant associated with the width, edge shape and the termination of the GNR[11]. Half-occupied interface states are predicted to occur at the heterojunctions between topologically trivial and nontrivial GNR segments (i.e., where the value of $Z_2$ changes across an interface as shown in Fig. 1a). These interface states, if aligned periodically in a superlattice, enable a hierarchy of topological quantum engineering since they are defined *locally* by topological phase discontinuities, while the superlattice's *global* electronic structure reflects the hybridization between them. The end properties of such a GNR superlattice are determined by the topology of the overall superlattice electronic structure. This scheme provides new strategies for modifying GNR bandgaps and even potentially inducing completely new GNR behaviors such as metallicity and magnetism out of individually semiconducting structural components.

Our strategy for topologically engineering new bottom-up GNR behavior relies on the synthesis of atomically-precise superlattices that are comprised of alternating topologically trivial 7-armchair GNR (AGNR) segments ($Z_2 = 0$) and topologically nontrivial 9-AGNR segments ($Z_2 = 1$) along the longitudinal GNR axis, thus leading to a 1D array of interface states[11]. If the coupling between these states is expressed as hopping amplitudes $t_1$ (for hopping across a 7-AGNR segment) and $t_2$ (for hopping across a 9-AGNR segment), then the band dispersion arising from the coupled topological interface states can be expressed in the standard two-band tight-binding form[15]

$$E_{\pm}(k) = \pm\sqrt{t_1^2 + t_2^2 + 2t_1 t_2 \cos(k)}, \qquad (1)$$

This leads to a tunable energy gap ($E_g = 2||t_1| - |t_2||$) and bandwidth ($W = |t_1| + |t_2| - E_g/2$) for the new bands that arise from purely topological considerations. The end properties of this superlattice, however, are not solely determined by $t_1$ and $t_2$, but must take into account the Zak or Berry phase of *all* the occupied π-electron bands[11]. By carefully controlling the atomic structure of the 7/9-AGNR superlattice termini, we ensure that the resulting global topological phase of the entire 7/9-AGNR superlattice is nontrivial. This mandates the existence of a series of end states in different energy gaps of this hierarchically-engineered 1D topological system.

The key to creating well-defined, periodic topological interface states is the design of molecular precursors that selectively link crystallographic unit cells of 7-AGNRs and 9-AGNRs into segments that have different topological phases. This is achieved by controlling the molecular structure of the 7/9-AGNR interface through careful design of building block **1** (Fig. 1b) (even small changes in the alignment of the interface structure can alter the topological phase of the constituent GNR segments[11]). The structural asymmetry between the two distinct reaction interfaces in **1**, a zig-zag edge on the side of the 7-AGNR and an armchair edge on the side of the 9-AGNR, leads to a sterically-enforced highly selective head-to-head/tail-to-tail polymerization during the on-surface synthesis (Figs. 1c, S1). Thermally induced cyclodehydrogenation of the resulting 7/9-polymer intermediates yields the topologically trivial/nontrivial superlattice of 7/9-AGNRs (Fig. 1c).

The synthesis of **1** is depicted in Fig 1b. Condensation of 6-bromo-(1,1'-biphenyl)-3-carbaldehyde with 2-naphthol yields xanthene **2** in 68% yield. Benzylic oxidation with lead(IV) oxide followed by dehydration of the intermediate xanthenol with tetrafluoroboric acid gives the pyrylium salt **3** in 69% over two steps. The molecular precursor **1** can be obtained as a minor product (<10% crude mixture, major product is the xanthenol) from the condensation of **3** with the sodium salt of 2-(10-bromoanthracen-9-yl)acetic acid. Analytically pure samples suitable for UHV deposition were obtained through multiple precipitations and recrystallizations from EtOH/CHCl$_3$. 7/9-AGNR superlattices were grown on Au(111) by sublimation of **1** onto a clean Au(111) single crystal under ultra-high vacuum (UHV) conditions. Fig. 1d shows an STM image of a sub-monolayer coverage of molecular precursor **1** on Au(111) at $T = 4$ K. The sample was subsequently annealed at 200 °C to induce the homolytic cleavage of C–Br bonds followed by radical step-growth polymerization to give *poly*-**1**. The polymer intermediate (Fig. 1e) exhibits a lattice periodicity that is twice the size of the molecular precursor **1**. This is consistent with the expectation that the lattice constant of the 7/9-AGNR supercell comprises a head-to-head molecular dimer. STM topography reveals a characteristic height profile and morphology that alternates between proto-9-AGNR and proto-7-AGNR segments (Fig. 1e). Annealing the sample at 300 °C induces cyclodehydrogenation and leads to the fully-fused 7/9-AGNR superlattices depicted in Figs. 1f,g. Bond-resolved STM (BRSTM) images of 7/9-AGNRs were acquired by recording the out-of-phase component of constant-current d$I$/d$V$ maps at low tip-sample bias using a functionalized STM tip (Fig. 1g). A representative BRSTM image depicted in Fig. 1g confirms the alternating sequence of short segments of 7-AGNRs and 9-AGNRs and the atomically precise 7/9 heterojunction interface characteristic for a 7/9-AGNR topological superlattice.

The local electronic structure of 7/9-AGNR superlattices was characterized using d$I$/d$V$ point spectroscopy as shown in Fig. 2a. All spectra were collected after calibrating the STM tip via the well-known Au(111) Shockley surface state. Spectra collected in the bulk of the 7/9-AGNR superlattice (at least 2.6 nm from a GNR end termination, corresponding to the length of one dimer unit) show a series of reproducible electronic states on both 7-AGNR and 9-AGNR segments, with peaks centered at $-1.14 \pm 0.07$ V (peak A), $-0.14 \pm 0.04$ V (peak B), $0.60 \pm 0.04$ V (peak C), and $1.61 \pm 0.04$ V (peak D). Since peaks B and C bracket the Fermi energy, $E_F$, our apparent experimental band gap for the 7/9-AGNR superlattice is $0.74 \pm 0.06$ eV. This gap is substantially smaller than the experimental band gaps measured by STS under similar conditions for both uniform 7-AGNRs (2.3 eV bandgap)[27] and 9-AGNRs (1.4 eV bandgap)[22] on Au(111). d$I$/d$V$ maps recorded at biases corresponding to the four peak energies in Fig. 2a reveal characteristic, reproducible patterns in the local density of states (LDOS) maps for each of these four bands (Fig. 2b).

As expected for a topologically nontrivial system in vacuum, spectroscopy performed near a terminal end of the 7/9-AGNR superlattice shows markedly different behavior compared to the bulk spectroscopy shown in Figs. 2a,b. Fig. 3b reveals three new spectral features confined to the last supercell of a 7/9-AGNR superlattice that are absent in the bulk (the new states are marked end states 1–3). The d$I$/d$V$ maps depicted in Fig. 3c show the characteristic LDOS patterns of end states 1–3 (by contrast, the d$I$/d$V$ maps of bulk features B and C show that they are absent from the last supercell). It is notable that end state 2 lies mid-gap between the bulk peaks B and C, while end states 1 and 3 lie within the A/B and C/D energy gaps, respectively. While the zigzag end termination shown in Fig. 3a is the most common 7/9-AGNR superlattice termination, the alternative termination (an armchair edge emerging from the 9-AGNR segment) was also observed and exhibits novel end state behavior as well (SI, Fig. S4).

The observed existence of end states 1–3 as well as the bulk behavior of the 7/9-AGNR superlattice follow the predictions of Ref. 11 since each of the two new interface-state-derived bands (B and C) can be shown to have a Zak phase equal to zero for the terminal geometry in Fig. 3a, making the system topologically nontrivial for all three gaps (A/B, B/C and C/D). To quantitatively verify the topological origins of the local electronic structure, we first compare the measurements to simulations performed using first-principles density functional theory (DFT) within the local-density approximation (LDA). Fig. 2d shows the theoretical bulk density of states (DOS) for a freestanding 7/9-AGNR superlattice. A series of peaks arise from the superlattice band structure (Fig. 4c) labeled valence band (VB), occupied topologically-induced band (OTB), unoccupied topologically-induced band (UTB), and conduction band (CB). The OTB and UTB are so named because they arise from the topologically protected interface states located at each internal 7/9-AGNR heterojunction. The relative positions of these four bands correlate with peaks A-D observed experimentally in the bulk region of a 7/9-AGNR superlattice as shown in Fig. 2a. Notably, the anomalously small band gap observed experimentally is nicely reproduced by the DFT calculations which predict a gap of 0.52 eV (see band structure in Fig. 4c). It is not surprising that this value is smaller than the gap observed experimentally (0.74 eV) since DFT tends to underestimate quasiparticle band gaps[17,28], even accounting for the screening effects of the underlying Au substrate[25]. Fig. 2c shows the theoretical LDOS maps at 4 Å above the plane of the 7/9-AGNR superlattice at energies corresponding to the VB, OTB, UTB, and CB. These LDOS maps are in excellent agreement with the experimental LDOS patterns shown in Fig. 2b. This agreement between *ab initio* theory and experiment confirms that peaks A–D observed in STS indeed originate from the intrinsic VB, OTB, UTB, and CB of the GNR superlattice.

The topological origin of the 7/9-AGNR superlattice bulk electronic properties is further evidenced by fitting Eq. (1) to the UTB and OTB band structure of Fig. 4c, which yields $t_1$ = 0.33 eV (for hopping across 9-AGNR segments) and $t_2$ = –0.07 eV (for hopping across 7-AGNR segments). The stronger hopping term across the 9-AGNR segment arises from its smaller intrinsic bandgap which allows the interface state to extend further into it and to overlap more strongly with adjacent states (Fig. 4a). This overlap causes the Bloch wavefunctions of the OTB and UTB at the Γ point to reflect bonding and anti-bonding interface states coupled through 9-AGNR segments (Fig. 4b). The presence of these topological interface-state-derived bands stands in stark contrast to the band structure of a nearly structurally equivalent, but topologically trivial, 7/9-AGNR superlattice (SI, Fig. S2) which completely lacks the two interface-state-derived bands due to the absence of variation in the value of $Z_2$ along its length. The substantial bandgap reduction seen in our 7/9-AGNR superlattice compared to the properties of individual 7-AGNRs and 9-AGNRs thus arises from the controlled incorporation of topological interface states into this bottom-up system.

The end state properties of the 7/9-AGNR superlattice are understood by examining the overall $Z_2$ value of the system for successive band occupation up to a particular bandgap. For the supercell associated with the experimentally observed end structure shown in Fig. 3a, the occupation of bands up to and including the VB results in the system being topologically nontrivial (i.e., $Z_2 = 1$ (SI, Fig. S3)), and thus requires the existence of a 7/9-AGNR/vacuum interface state in the VB/OTB energy gap (i.e., the experimental state labeled end state 1 in Fig. 3b). The behavior in the next OTB/UTB energy gap is determined by the Zak phase of the OTB plus those of the entire band complex below it. Although the OTB and UTB arise directly from coupled topological interface states, analysis of the Zak phase of these bands shows that it is zero (topologically trivial) for each band for the terminating geometry considered (SI, Fig. S3). The overall value of $Z_2$ thus remains $Z_2 = 1$ for the OTB/UTB and UTB/CB gaps, making the existence of topologically-protected 7/9-AGNR/vacuum end states required in both energy gaps, just as seen experimentally (i.e., end states 2 and 3 in Fig. 3b). (Similar analysis reveals nontrivial topology for the other, less common experimentally observed superlattice end structure (SI, Fig. S4)).

This topological behavior can also be clearly seen in our simulations of the end region of a 7/9-AGNR superlattice calculated for a finite 7/9-AGNR consisting of eight supercells. The LDOS of this structure exactly reproduces end states 1-3 in the three energy gaps, as discussed above (SI, Fig. S3). A direct comparison between the experimental d$I$/d$V$ maps and the calculated LDOS maps of end states 1–3 shows excellent agreement between theory and experiment (Fig. 3c). Similarly, the experimental d$I$/d$V$ maps and the calculated LDOS maps of the OTB and UTB show high intensity throughout the bulk 7/9-AGNR superlattice but decay rapidly in the last supercell that terminates the GNR.

In conclusion, we have demonstrated that it is possible to rationally engineer both the local and global GNR electronic topology via careful design of molecular precursors used in bottom-up synthesis. This novel approach enables the deterministic design of topological interface states both in the GNR bulk as well as in the GNR/vacuum termination region. Superlattices of topological interface states allow formation of new bulk frontier bands (the OTB and UTB) that are energetically distinct from the intrinsic band structures associated with the parent 7- and 9-AGNRs. In principle, the properties of these new, topologically-induced bands can be fine-tuned through topology-conserving modifications of the superlattice components to create effective AFM Heisenberg spin-1/2 chains with robust spin centers at each internal 7/9-AGNR interface[11]. If placed in close proximity to a superconductor, the ends of these AFM chains are predicted to host Majorana fermion states[29].

## Methods

**Precursor Synthesis and GNR Superlattice Growth.** Full details of the synthesis and characterization of **1**–**4** are given in the Supplementary Information. 7/9-AGNR superlattices were grown on a Au(111) single crystal under UHV conditions. Atomically clean Au(111) surfaces were prepared through iterative Ar$^+$ sputter/anneal cycles. Sub-monolayer coverage of **1** on atomically clean Au(111) was obtained by sublimation using a home-built Knudsen cell evaporator for 20–30 min at crucible temperatures of 200–215 °C. After deposition the surface temperature was slowly ramped ($\leq$ 2 K min$^{-1}$) to 200 °C and held at this temperature for 30 min to induce the radical-step growth polymerization, then ramped slowly ($\leq$ 2 K min$^{-1}$) to 300 °C and held there for 30 min to induce cyclodehydrogenation.

**STM Measurements.** All STM experiments were performed using a commercial Createc LT-STM held at $T \approx$ 4K using PtIr STM tips. d$I$/d$V$ measurements were recorded using a lock-in amplifier with a modulation frequency of 581 Hz and a modulation amplitude of $V_{RMS}$ = 10–20 mV. d$I$/d$V$ point spectra were recorded under open feedback loop conditions. d$I$/d$V$ maps were collected under constant current conditions. BRSTM images were obtained by mapping the out-of-phase d$I$/d$V$ signal collected during a low bias (20 mV) d$I$/d$V$ map. Peak positions in d$I$/d$V$ point spectroscopy were determined by fitting the spectra with Lorentzian peaks. Each peak position is based on an average of approximately 80 spectra collected on 15 GNRs with 17 different tips, all of which were first calibrated to the Au(111) Shockley surface state.

**Calculations.** First-principles calculations of GNR superlattices were performed using DFT in the LDA as implemented in the Quantum Espresso package[30,31]. A supercell arrangement was used with vacuum regions carefully tested to avoid interactions between the superlattice and its periodic image. We used norm-conserving pseudopotentials with a planewave energy cut-off of 60 Ry. The structure was fully relaxed until the force on each atom was smaller than 0.02 eV Å$^{-1}$. All σ dangling bonds on the edges and the ends of the GNR were capped by hydrogen atoms. A Gaussian broadening of 0.05 eV was used in the evaluation of DOS.

6    Xu, C. & Moore, J. E. Stability of the quantum spin Hall effect: Effects of interactions, disorder, and $Z_2$ topology. *Phys. Rev. B* **73**, 045322 (2006).

7    Hsieh, D. *et al.* A topological Dirac insulator in a quantum spin Hall phase. *Nature* **452**, 970 (2008).

8    Hsieh, D. *et al.* Observation of Unconventional Quantum Spin Textures in Topological Insulators. *Science* **323**, 919 (2009).

9    Teo, J. C. Y., Fu, L. & Kane, C. L. Surface states and topological invariants in three-dimensional topological insulators: Application to $Bi_{1-x}Sb_x$. *Phys. Rev. B* **78**, 045426 (2008).

10   Nishide, A. *et al.* Direct mapping of the spin-filtered surface bands of a three-dimensional quantum spin Hall insulator. *Phys. Rev. B* **81**, 041309 (2010).

11   Cao, T., Zhao, F. & Louie, S. G. Topological Phases in Graphene Nanoribbons: Junction States, Spin Centers, and Quantum Spin Chains. *Phys. Rev. Lett.* **119**, 076401 (2017).

12   Cheon, S., Kim, T.-H., Lee, S.-H. & Yeom, H. W. Chiral solitons in a coupled double Peierls chain. *Science* **350**, 182 (2015).

13   Kim, T.-H., Cheon, S. & Yeom, H. W. Switching chiral solitons for algebraic operation of topological quaternary digits. *Nat. Phys.* **13**, 444 (2017).

14   Brazovskii, S., Brun, C., Wang, Z.-Z. & Monceau, P. Scanning-Tunneling Microscope Imaging of Single-Electron Solitons in a Material with Incommensurate Charge-Density Waves. *Phys. Rev. Lett.* **108**, 096801 (2012).

15   Su, W. P., Schrieffer, J. R. & Heeger, A. J. Solitons in Polyacetylene. *Phys. Rev. Lett.* **42**, 1698-1701 (1979).
12

**Supplementary Information**

Supplementary information contains additional details regarding synthesis and characterization of precursor **1**, STS, DFT, and topological characterization of GNR superlattices.

**Acknowledgements**

Research supported by the Office of Naval Research MURI Program N00014-16-1-2921 (precursor design, STM spectroscopy, band structure), by the U.S. Department of Energy (DOE), Office of Science, Basic Energy Sciences (BES) under award no. DE-SC0010409 (precursor synthesis and characterization) and the Nanomachine Program award no. DE-AC02-05CH11231 (surface growth, heterojunction analysis), by the Center for Energy Efficient Electronics Science NSF Award 0939514 (end state modelling), and by the National Science Foundation under grant DMR-1508412 (development of theory formalism). Computational resources have been provided

**Author Information**

The Reprints and permissions information is available at www.nature.com/reprints. The authors declare no competing financial interests. Correspondence and requests for materials should be addressed to ffischer@berkeley.edu, crommie@berkeley.edu, or sglouie@berkeley.edu.

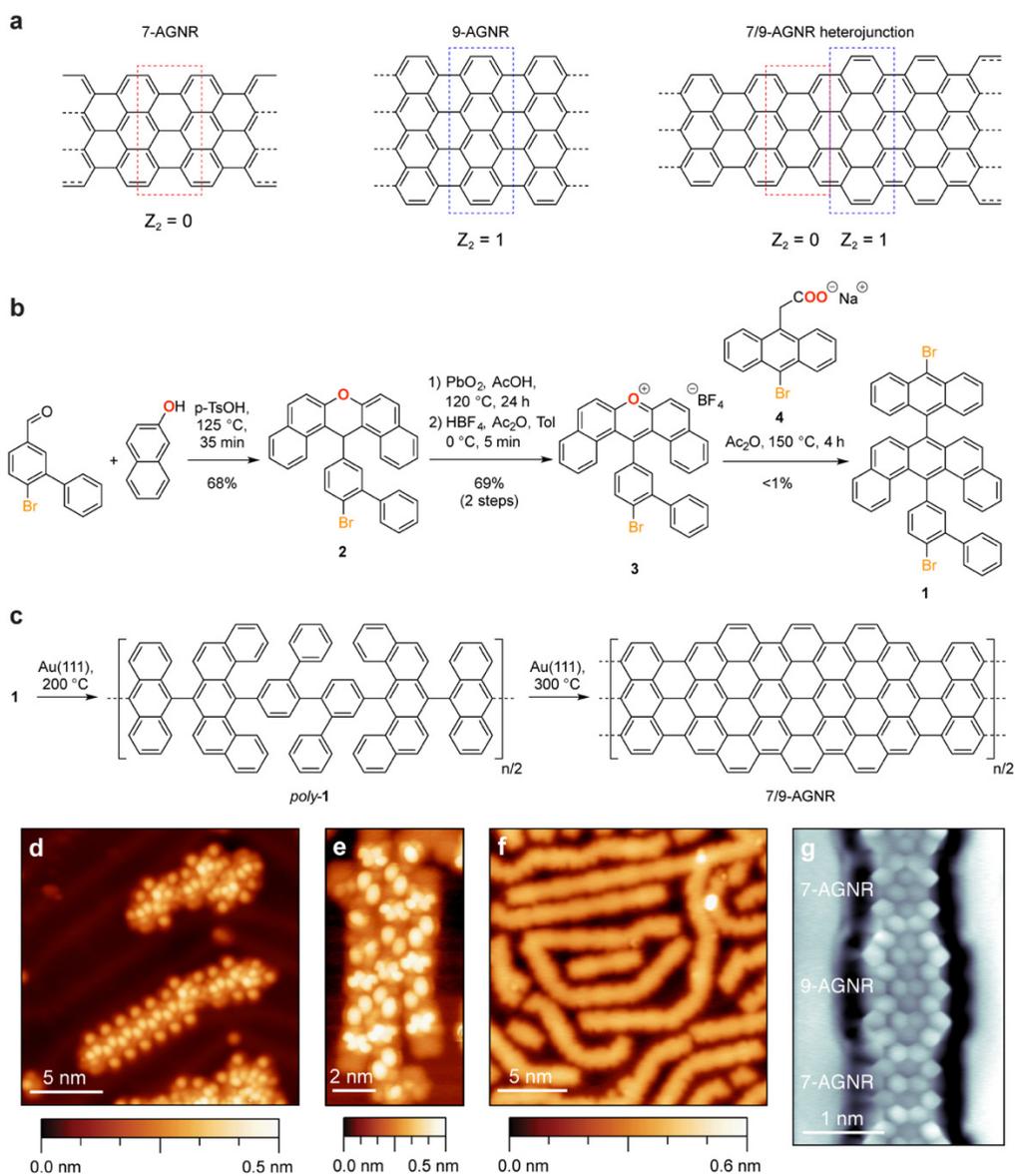

**Figure 1 | Bottom-up Synthesis of 7/9-AGNR superlattice on Au(111). a,** Schematic representation of the $Z_2$ invariant associated with particular terminations of 7- and 9-AGNRs and at the interface of a 7/9-AGNR heterojunction. **b,** Synthesis of molecular precursor **1**. **c,** Schematic representation of the stepwise thermally-induced on-surface growth of a 7/9-AGNR superlattice from molecular precursor **1**. *Poly*-**1** is formed upon annealing to 200 °C and full cyclization occurs at 300 °C. Activated precursors polymerize in a head-to-head orientation due

to the distinctive steric constraints of the two active sites. **d,** STM topography of precursor **1** as deposited on Au(111) ($V_s$ = 1.00 V, $I_t$ = 30 pA). **e,** Polymer island on Au(111) after annealing to 200 °C ($V_s$ = 1.00 V, $I_t$ = 30 pA). **f,** Fully cyclized 7/9-AGNR superlattice on Au(111) after annealing to 300 °C ($V_s$ = 0.20 V, $I_t$ = 30 pA). **g,** BRSTM image of 7/9-AGNR superlattice shows the bond resolved structure of the heterojunction interface ($V_s$ = 0.02 V, $I_t$ = 80 pA, $f$ = 581 Hz, $V_{ac}$ = 12 mV). All STM data obtained at $T$ = 4 K.

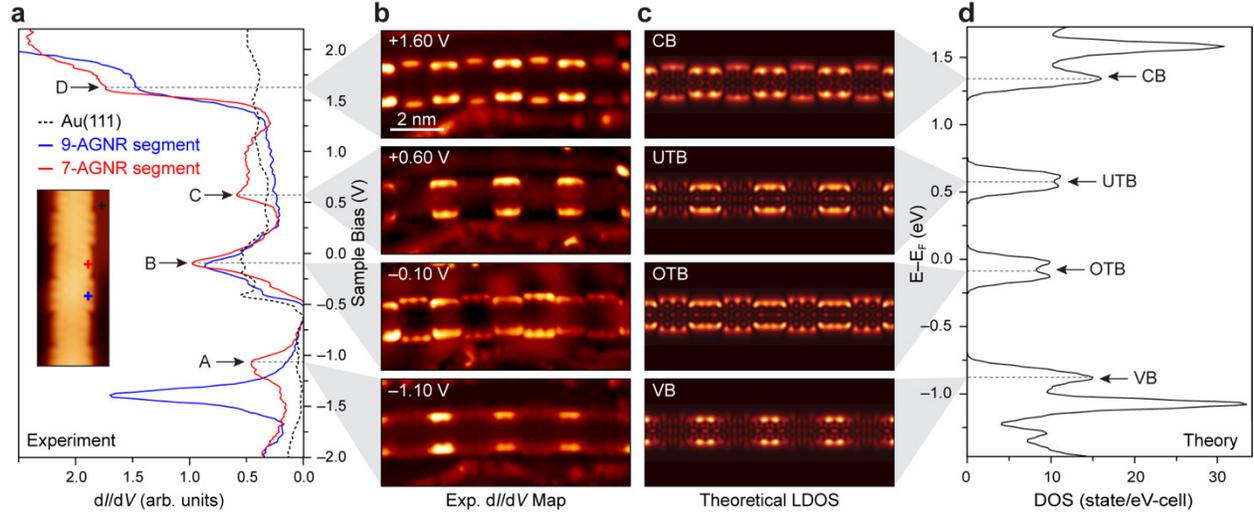

**Figure 2 | Electronic structure of 7/9-AGNR superlattice. a,** Inset: STM topography of 7/9-AGNR superlattice ($V_s = -0.10$ V, $I_t = 80$ pA). Red (blue) curve shows $dI/dV$ point spectroscopy collected on 7-AGNR (9-AGNR) segment (tip position marked by plus sign). Dotted black curve shows spectrum collected on bare Au(111). Spectroscopy parameters: $V_{ac} = 10$ mV, $f = 581$ Hz. **b,** Constant-current $dI/dV$ maps obtained at voltages corresponding to the A, B, C, and D peaks marked by arrows in **a** ($I_t = 100$ pA, $V_{ac} = 10$ mV, $f = 581$ Hz). **c,** Local density of states maps calculated via DFT at the energies of the 7/9-AGNR superlattice VB, OTB, UTB, and CB marked by arrows in **d** (obtained 4 Å above the GNR plane). **d,** Calculated density of states obtained via DFT for the 7/9-AGNR superlattice (0.05 eV Gaussian broadening used). All STM data obtained at $T = 4$K.

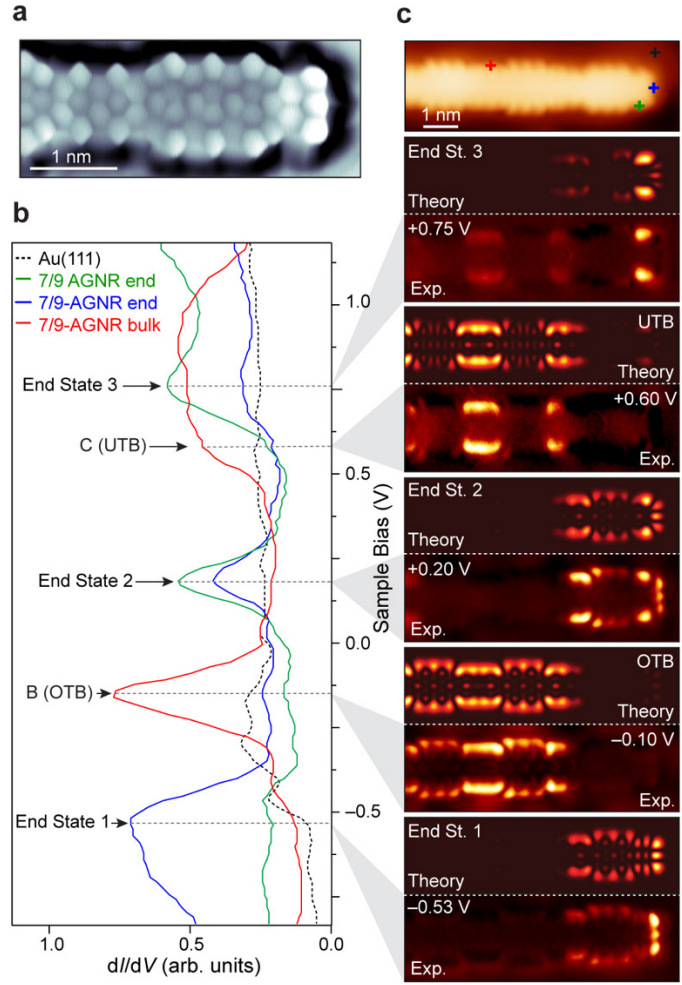

**Figure 3 | Electronic structure of 7/9-AGNR superlattice end states. a,** BRSTM image of most common 7/9-AGNR superlattice end termination ($V_s$ = 0.02 V, $I_t$ = 80 pA, $f$ = 581 Hz, $V_{ac}$ = 12 mV). **b,** Red (green, blue) curve shows d$I$/d$V$ point spectroscopy collected above the 7/9-AGNR superlattice bulk (end). Dotted black curve shows spectrum collected on bare Au(111). Spectroscopy parameters: $V_{ac}$ = 20 mV, $f$ = 581 Hz. **c,** STM topography showing the tip position for STS in **b** (marked by a plus sign) ($V_s$ = –0.10 V, $I_t$ = 80 pA). Experimental d$I$/d$V$ map (bottom) compared to corresponding theoretical LDOS maps (top) for end state 1, OTB, end state 2, UTB, and end state 3. d$I$/d$V$ map parameters: $I_t$ = 50 pA, $V_{ac}$ = 20 mV, $f$ = 581 Hz. All

STM data obtained at $T$ = 4 K. DFT calculated LDOS map at fixed energy simulated at a height of 4 Å above freestanding 7/9-AGNR superlattice comprised of eight supercells (see SI).

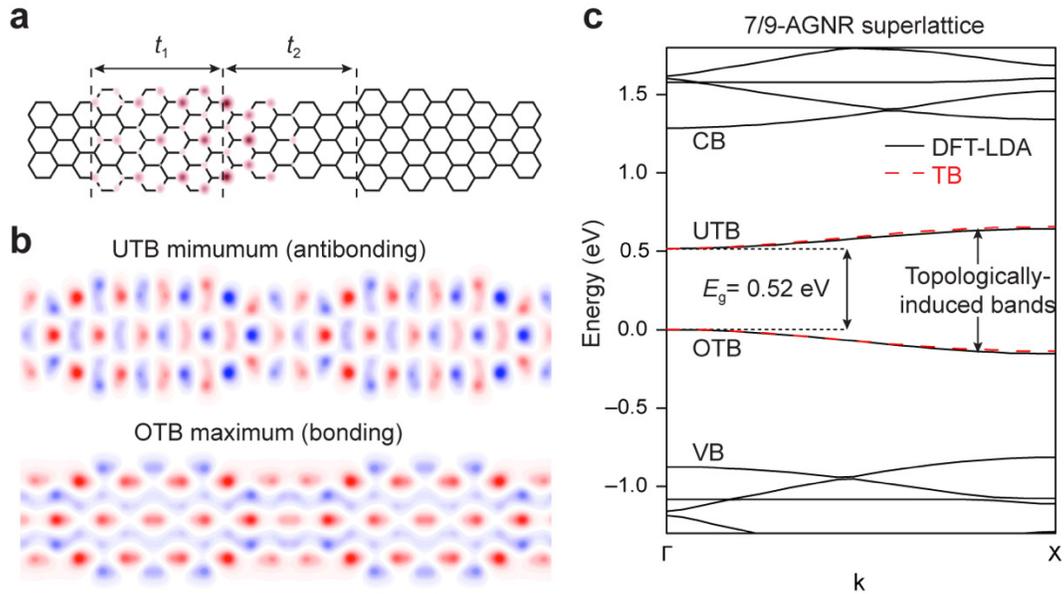

**Figure 4 | Topological bands in 7/9-AGNR superlattice. a,** Wireframe sketch of the 7/9-AGNR superlattice with superimposed charge density for only a single, isolated interface state. Arrows represent hopping amplitudes that couple interface states through different AGNR segments. **b,** DFT-LDA wavefunctions for the OTB maximum and UTB minimum are composed of bonding and antibonding linear combinations of adjacent interface-state wavefunctions. The wavefunctions are plotted in the plane 1Å above the GNR atomic plane in order to demonstrate bonding symmetry. **c,** Solid black curves show the DFT-LDA band structure of a freestanding 7/9-AGNR superlattice. Dashed red curve shows a tight-binding fit to DFT OTB and UTB using equation (1).

# Supplementary Information:

# Topological Band Engineering of Graphene Nanoribbons

**Daniel J. Rizzo[1,†], Gregory Veber[2,†], Ting Cao[1,3,†], Christopher Bronner[1], Ting Chen[1], Fangzhou Zhao[1], Henry Rodriguez[1], Steven G. Louie[1,3,*], Michael F. Crommie[1,3,4,*], Felix R. Fischer[2,3,4,*]**

[1]Department of Physics, University of California, Berkeley, CA 94720, USA. [2]Department of Chemistry, University of California, Berkeley, CA 94720, USA. [3]Materials Sciences Division, Lawrence Berkeley National Laboratory, Berkeley, CA 94720, USA. [4]Kavli Energy NanoSciences Institute at the University of California Berkeley and the Lawrence Berkeley National Laboratory, Berkeley, California 94720, USA.
† These authors contributed equally to this work.
* Corresponding authors

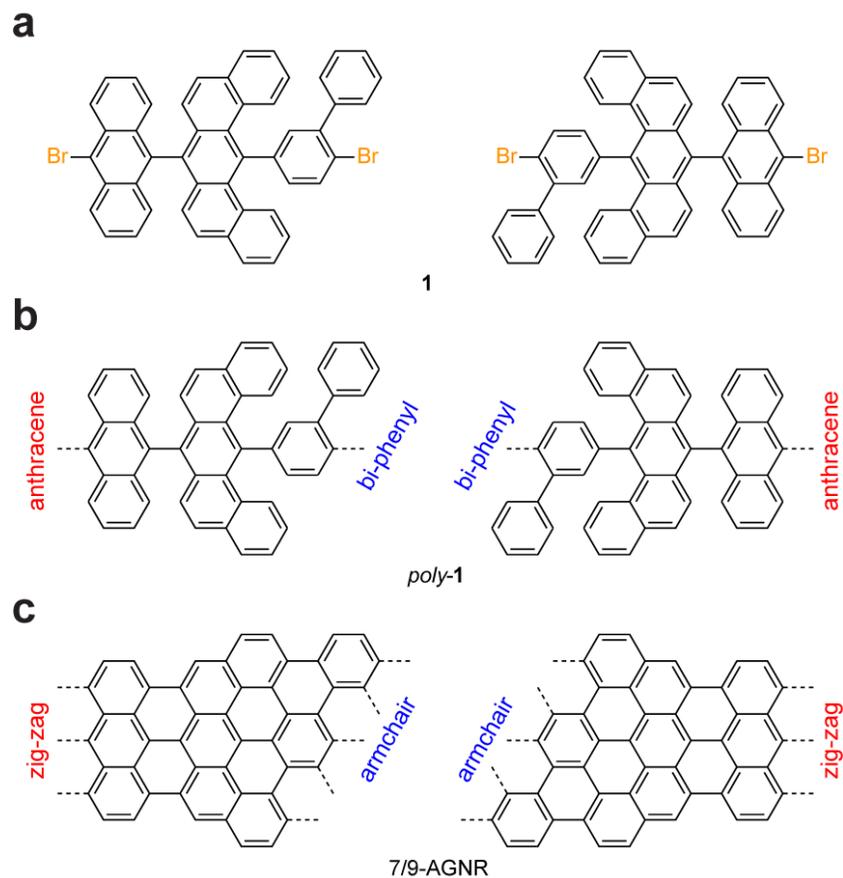

**Supplementary Figure S1 | Sterically enforced site selective polymerization. a,** molecular precursor **1**. **b,** sterically distinct reaction sites during radical chain growth polymerization (anthracene vs. biphenyl). **c,** the corresponding edge structures in the fully formed GNR (armchair and zig-zag termination, respectively)

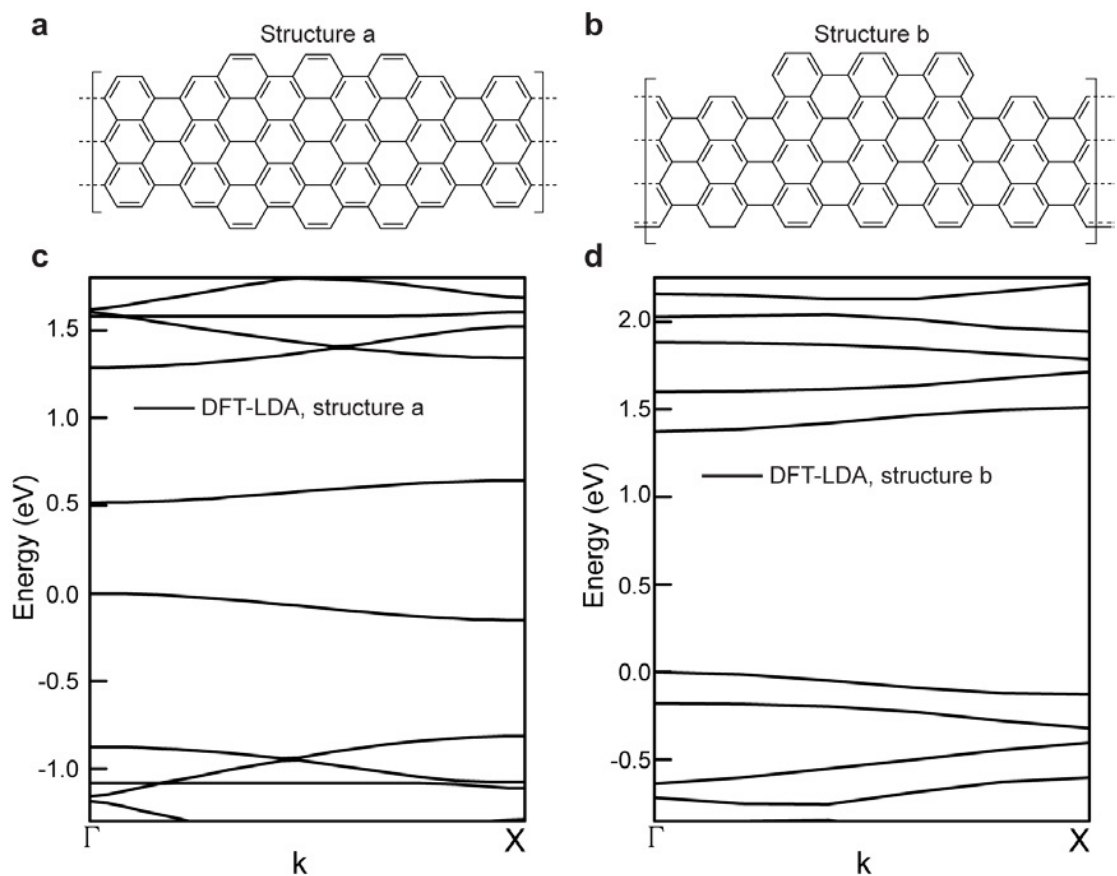

**Supplementary Figure S2 | Electronic structure of two different 7/9-AGNR superlattices. a,** Unit cell of structure a and **c,** DFT calculated band structure for 7/9-AGNR superlattice composed of topologically nontrivial interfaces shows two new topologically-induced bands. **b,** Unit cell of structure b and **d,** DFT-calculated band structure of 7/9-AGNR superlattice - composed of topologically trivial interfaces showing no topologically-induced bands in the energy gap region.

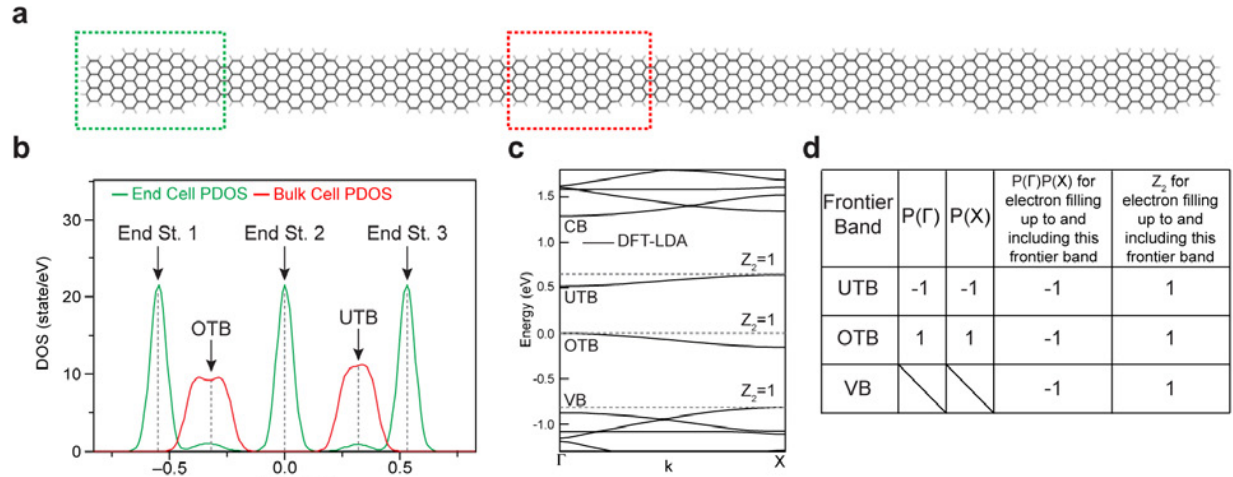

**Supplementary Figure S3 | Electronic structure of finite nontrivial 7/9-AGNR superlattice.**
**a,** Fully-relaxed finite (8-supercell) 7/9-AGNR superlattice with end (green) and bulk (red) unit cells indicated. **b,** DFT calculated PDOS of finite (8-supercell) 7/9-AGNR superlattice obtained from the end unit cell (green) and a bulk unit cell (red) (gaussian broadening of 0.05 eV was used here). Three end states are seen that closely correspond to the experimental end states shown in Fig. 3b of the manuscript. **c,** DFT-calculated band structure of 7/9-AGNR showing the overall value of $Z_2$ for occupation up to all three energy gaps around $E_F$ based on the edge structure shown in Fig. 3a of the manuscript. **d,** Chart of frontier band parity eigenvalues and corresponding $Z_2$ invariants for electron filling up to and including a given frontier band.

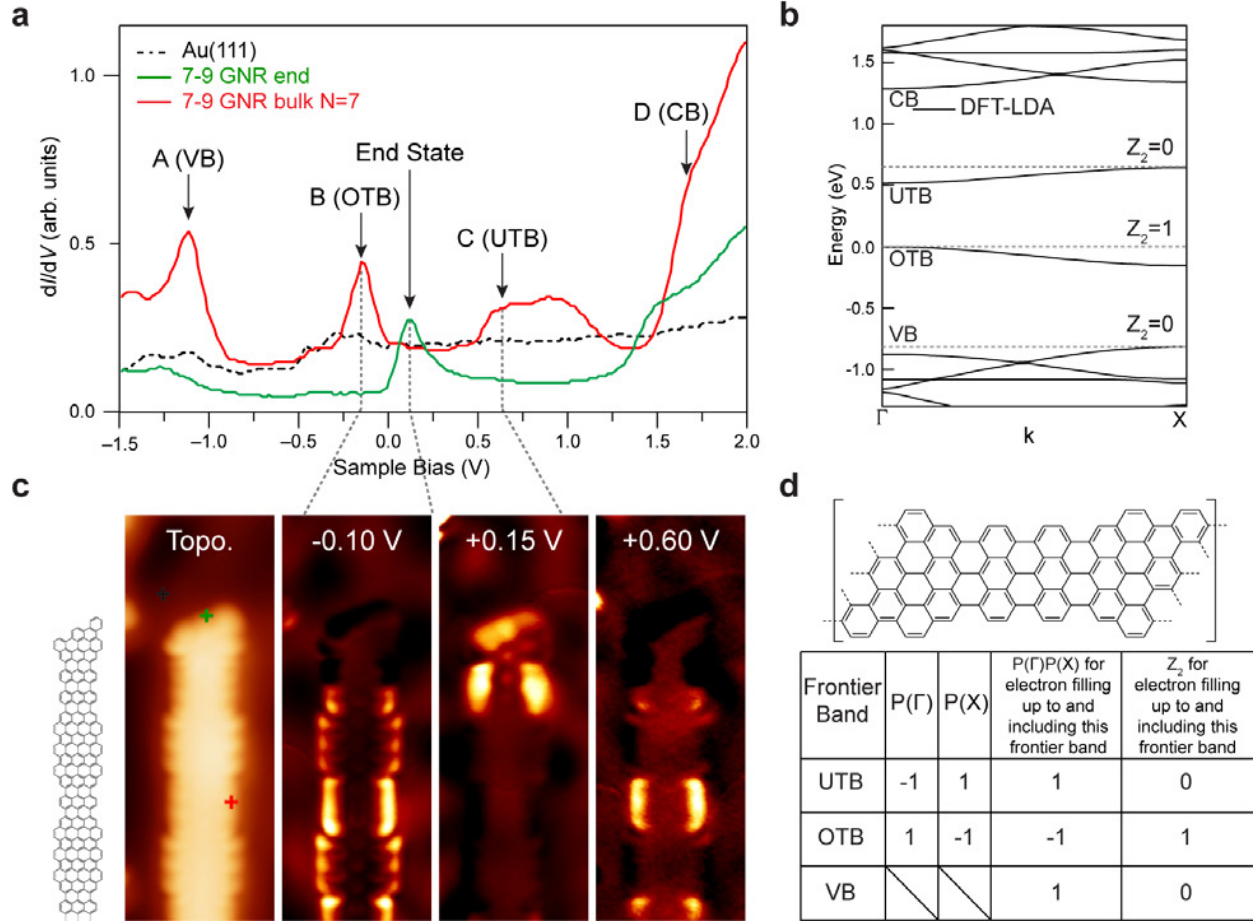

**Supplementary Figure S4 | Topologically nontrivial 7/9-AGNR with different end structure. a,** Red (green) curve shows d$I$/d$V$ point spectroscopy collected on 7/9-AGNR superlattice bulk (end) region. Dashed black curve shows spectrum collected on bare Au(111). Only one end state is observed that lies in the energy gap between the OTB and UTB. Spectroscopy parameters: $V_{ac}$ = 20 mV, $f$ = 581 Hz. **b,** DFT-calculated band structure of 7/9-AGNR shows an overall topologically nontrivial phase for the edge structure shown in **c** only for bands filled up to and including the OTB. **c,** Sketch of GNR structure and STM topographic image of additional 7/9-AGNR end terminus that is seen for <10% of all 7/9 AGNR superlattices in experiment. Experimental topography and d$I$/d$V$ maps are shown for the OTB, the end state, and the UTB (topography: $V_s$ = –0.10 V, $I_t$ = 50 pA; $dI/dV$ maps: $I_t$ = 50 pA, $f$ = 581 Hz, $V_{ac}$ = 20

mV). **d,** Unit cell commensurate with uncommon end terminus shown in **c**, along with chart of frontier band parity eigenvalues and corresponding $Z_2$ invariants for electron filling up to and including a given frontier band. For this edge structure only the OTB/UTB gap is topologically nontrivial and supports a topologically protected end state. The UTB/CB and VB/OTB gaps are topologically trivial and do not support end states, unlike the behavior seen for the other, more common termination shown in Fig. 3a of the main manuscript.

**Materials and General Methods.** Unless otherwise stated, all manipulations of air and/or moisture sensitive compounds were carried out in oven-dried glassware, under an atmosphere of Argon. All solvents and reagents were purchased from Alfa Aesar, Spectrum Chemicals, Acros Organics, TCI America, and Sigma-Aldrich and were used as received unless otherwise noted. Organic solvents were dried by passing through a column of alumina and were degassed by vigorous bubbling of N2 or Ar through the solvent for 20 min. Flash column chromatography was performed on SiliCycle silica gel (particle size 40–63 μm). Thin layer chromatography was carried out using SiliCycle silica gel 60 Å F-254 precoated plates (0.25 mm thick) and visualized by UV absorption. All $^1$H and $^{13}$C NMR spectra were recorded on Bruker AV-300, AVB-400, AV-600, DRX-500, and AV-500 MHz spectrometers, and are referenced to residual solvent peaks (CDCl$_3$ $^1$H NMR = 7.26 ppm, $^{13}$C NMR = 77.16 ppm; CD$_2$Cl$_2$ $^1$H NMR = 5.32 ppm, $^{13}$C NMR = 53.84 ppm; [D$_6$]DMSO $^1$H NMR = 2.50 ppm, $^{13}$C NMR = 39.52 ppm; CD$_3$OD $^1$H NMR = 3.31 ppm, $^{13}$C NMR = 49.00 ppm; C$_2$D$_2$Cl$_4$ $^1$H NMR = 6.00 ppm, $^{13}$C NMR = 73.78 ppm ). ESI mass spectrometry was performed on a Finnigan LTQFT (Thermo) spectrometer in positive ionization mode. MALDI mass spectrometry was performed on a Voyager-DE PRO (Applied Biosystems Voyager System 6322) in positive mode using a matrix of dithranol. 2-(anthracen-9-yl)acetic acid was synthesized according to literature procedures[1].

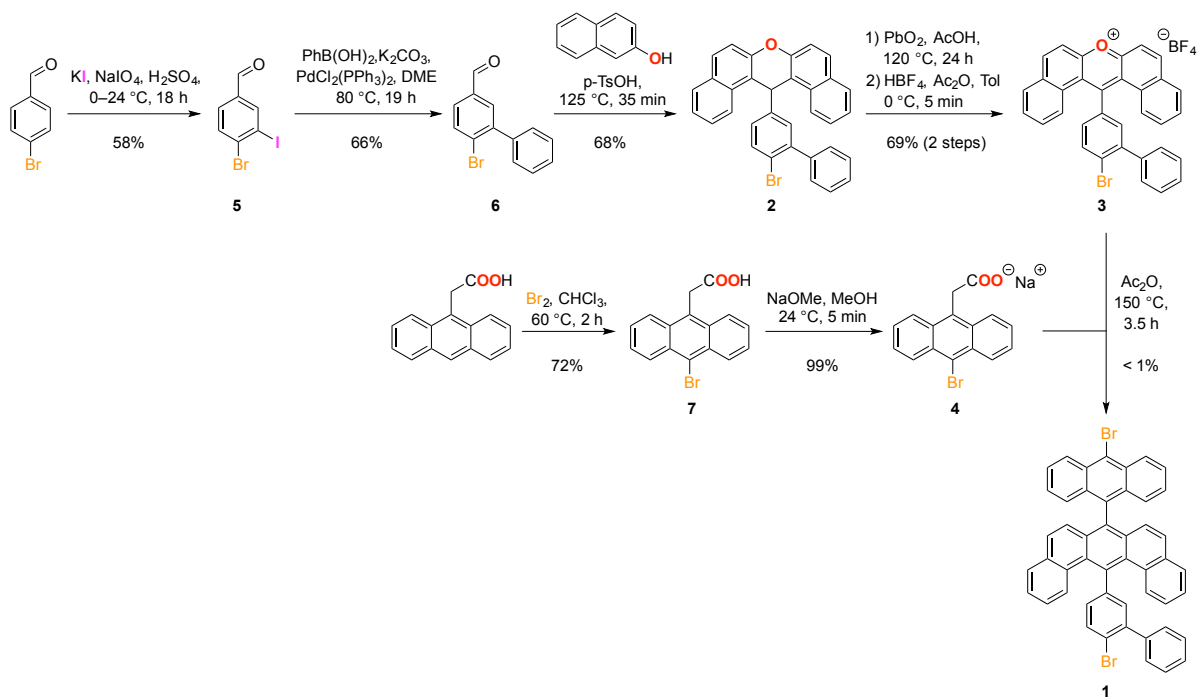

**Supplementary Scheme S1 | Synthesis of molecular precursor 1.**

**4-bromo-3-iodobenzaldehyde (5)** A 250 mL three neck flask was charged with powered KI (1.21 g, 7.30 mmol) and NaIO$_4$ (518 mg, 2.43 mmol) in 90% H$_2$SO$_4$ at 0 °C. The purple solution was warmed to 24 °C over the period of 30 min, after which 4-bromo-benzaldehyde (1.5 g, 8.11 mmol) was added in one portion. The purple solution was stirred at 24 °C for 18 h. The solution was added to ice and the precipitate was filtered. The solid was dissolved in EtOAc and washed three times with saturated sodium bisulfite, once with saturated aqueous NaCl, dried over MgSO$_4$ and concentrated on a rotary evaporator. Recrystallization from EtOH yielded **5** (1.463 g, 4.71 mmol, 58%) as a colorless solid. $^1$H NMR (400 MHz, CDCl$_3$, 22 °C) δ = 9.90 (s, 1H), 8.33 (d, $J$ = 1.9, 1H), 7.81 (d, $J$ = 8.2, 1H), 7.70 (dd, $J$ = 8.2, 1.9, 1H) ppm; $^{13}$C {$^1$H} NMR (101 MHz, CDCl$_3$, 22 °C) δ = 189.8, 141.4, 137.1, 136.2, 133.5, 129.8, 102.1 ppm; HRMS (EI-TOF) m/z: [C$_7$H$_4$BrIO]$^+$ calcd. [C$_7$H$_4$BrIO] 311.8470; found 311.8469.

**6-bromo-[1,1'-biphenyl]-3-carbaldehyde (6)** A 100 mL Schlenk flask was charged with **5** (1.80 g, 5.78 mmol), phenyl boronic acid (740 mg, 6.07 mmol) and $K_2CO_3$ (2.00 g, 14.5 mmol) in DME (21 mL) and $H_2O$ (2.3 mL). The suspension was degassed via $N_2$ sparging for 15 min, after which $PdCl_2(PPh_3)_2$ (60.7 mg, 0.087 mmol) was added. The suspension was stirred at 80 °C for 19 h under $N_2$. The reaction mixture was cooled to 24 °C and the solvent was removed via rotary evaporator. The suspension was added to $H_2O$ and the aqueous phase was washed three times with $Et_2O$. The combined organic phases were washed with saturated aqueous NaCl, dried over $MgSO_4$ and concentrated on a rotary evaporator. Recrystallization from EtOH yielded **6** (0.989 g, 3.79 mmol, 66 %) as colorless prism shaped crystals. $^1$H NMR (400 MHz, $CD_2Cl_2$, 22 °C) δ = 10.01 (s, 1H), 7.87 (d, $J$ = 8.2, 1H), 7.81 (d, $J$ = 2.1, 1H), 7.71 (dd, $J$ = 8.2, 2.1, 1H), 7.51–7.40 (m, 5H) ppm; $^{13}$C {$^1$H} NMR (151 MHz, $CDCl_3$, 22 °C) δ = 191.29, 143.7, 139.9, 135.6, 134.2, 132.4, 130.0, 129.4, 129.1, 128.4 ppm; HRMS (EI-TOF) m/z: $[C_{13}H_9OBr]^+$ calcd. $[C_{13}H_9OBr]$ 259.9837; found 259.9830.

**14-(6-bromo-[1,1'-biphenyl]-3-yl)-14H-dibenzo[a,j]xanthene (2)** A 50 mL three neck flask was charged with **6** (997.6 mg, 3.82 mmol), 2-naphthol (1.101 g, 7.64 mmol), and *p*-TSA monohydrate (14.6 mg, 0.076 mmol). The flask was evacuated and backfilled with $N_2$ twice then heated to 125 °C for 35 min. The solid was cooled to 24 °C and stirred in a mixture of $EtOH/H_2O$ (1/3) for 1 h. The solid was filtered and recrystallized from $EtOH/CHCl_3$ to yield **2** (1.35 g, 2.63 mmol, 68 %) as a colorless solid. $^1$H NMR (500 MHz, $CD_2Cl_2$, 22 °C) δ = 8.38 (d, $J$ = 8.5, 2H), 7.88 (d, $J$ = 8.1, 2H), 7.84 (d, $J$ = 8.9, 2H), 7.62 (ddd, $J$ = 8.4, 6.8, 1.4, 2H), 7.56 (d, $J$ = 2.4, 1H), 7.51–7.41 (m, 5H), 7.41–7.32 (m, 4H), 7.27–7.22 (m, 2H), 6.53 (s, 1H) ppm; $^{13}$C {$^1$H} NMR (151 MHz, $CDCl_3$, 22 °C) δ = 148.9, 144.6, 142.3, 141.1, 133.5, 131.4, 131.3, 130.9, 129.6, 129.3, 129.1, 128.8, 128.1,

127.7, 127.0, 124.5, 122.6, 120.6, 118.2, 116.7, 37.7 ppm; HRMS (EI-TOF) m/z: $[C_{33}H_{21}BrO]^+$ calcd. $[C_{33}H_{21}BrO]$ 514.0755; found 514.0759.

**14-(6-bromo-[1,1'-biphenyl]-3-yl)dibenzo[a,j]xanthen-7-ium tetrafluoroborate (3)** A 100 mL two neck flask was charged with **2** (1.35 g, 2.63 mmol) and $PbO_2$ (1.26 g, 5.26 mmol) in AcOH (35 mL). The reaction mixture was stirred at 120 °C for 24 h under $N_2$. The reaction mixture was cooled to 24 °C and added to ice. The solid was filtered, washed with $H_2O$, dried under vacuum at 80 °C for 18 h. The crude 14-(6-bromo-[1,1'-biphenyl]-3-yl)-14H-dibenzo[a,j]xanthen-14-ol (1.38 g) was used without further purification. A 25 mL Schlenk flask was charged with 14-(6-bromo-[1,1'-biphenyl]-3-yl)-14H-dibenzo[a,j]xanthen-14-ol (1.38 g, 2.61 mmol) in $Ac_2O$ (7.8 mL) and anhydrous toluene (2.6 mL) under $N_2$. The brown suspension was cooled to 0 °C and $HBF_4$ (1.3 mL, 10.56 mmol, 48 wt.% in $H_2O$) was added dropwise. At 0 °C anhydrous $Et_2O$ was added, the orange precipitate was filtered and washed with anhydrous $Et_2O$ yielding **3** (1.082 g, 1.81 mmol, 69 %) as an orange solid. $^1H$ NMR (500 MHz, $CD_2Cl_2$, 22 °C) δ = 8.84 (d, $J$ = 9.1, 2H), 8.31 (d, $J$ = 9.1, 2H), 8.27–8.23 (m, 3H), 7.87 (t, $J$ = 7.5, 2H), 7.64 (ddd, $J$ = 8.7, 7.1, 1.5, 2H), 7.52 (d, $J$ = 2.3, 1H), 7.49–7.44 (m, 3H), 7.43–7.40 (m, 5H) ppm; $^{13}C$ {$^1H$} NMR (101 MHz, $C_2D_2Cl_4$, 22 °C) δ = 164.6, 159.2, 147.4, 147.0, 138.7, 137.3, 137.0, 132.6, 131.9, 130.7, 130.1, 128.9, 128.8, 128.7, 128.5, 128.4, 127.8, 126.6, 126.6, 121.6, 117.1 ppm; HRMS (ESI-TOF) m/z: $[C_{33}H_{20}BrO]^+$ calcd. $[C_{33}H_{20}BrO]$ 511.0692; found 511.0687.

**2-(10-bromoanthracen-9-yl)acetic acid (7)** A 50 mL two neck flask was charged with 2-(anthracen-9-yl)acetic acid (407 mg, 1.72 mmol) in $CHCl_3$ (17 mL). A solution of $Br_2$ (300.2 mg, 1.89 mmol) in $CHCl_3$ (2 mL) was added dropwise at 24 °C over a period of 15 min. The reaction mixture was heated to 60 °C for 2 h then cooled to 0 °C. The precipitate was filtered and washed with cold $CH_2Cl_2$ yielding **7** (390.2 mg, 1.24 mmol, 72 %) as a slightly yellow solid. $^1H$ NMR

(500 MHz, DMSO-$d_6$, 22 °C) δ = 8.51 (d, $J$ = 8.7, 2H), 8.39 (d, $J$ = 8.7, 2H), 7.75–7.71 (m, 2H), 7.69–7.65 (m, 2H), 4.69 (s, 2H) ppm; $^{13}$C {$^1$H} NMR (101 MHz, DMSO-$d_6$, 22 °C) δ = 172.4, 130.9, 129.7, 129.3, 127.7, 127.7, 126.6, 125.5, 121.8, 33.9 ppm; HRMS (EI-TOF) m/z: [$C_{16}H_{11}BrO_2$]$^+$ calcd. [$C_{16}H_{11}BrO_2$] 315.9922; found 315.9923.

**sodium 2-(10-bromoanthracen-9-yl)acetate (4)** A 50 mL round bottom flask was charged with **7** (337.9 mg, 1.07 mmol) in MeOH (6 mL). The reaction mixture was stirred at 24 °C and NaOMe (2.14 mL, 1.07 mmol, 0.5 M in MeOH) was added dropwise under $N_2$. The solution was stirred for 5 min and the solvent was removed via rotary evaporator yielding **4** (361 mg, quant.) as a yellow solid that was used directly in the next step. $^1$H NMR (400 MHz, CD$_3$OD, 22 °C) δ = 8.58–8.53 (m, 2H), 8.47–8.42 (m, 2H), 7.63–7.53 (m, 4H), 4.55 (s, 2H) ppm.

**14-(6-bromo-[1,1'-biphenyl]-3-yl)-7-(10-bromoanthracen-9-yl)benzo[m]tetraphene (1)** An oven dried 50 mL Schlenk flask with reflux condenser was charged with **3** (587.1 mg, 0.98 mmol) and **4** (991 mg, 2.94 mmol) in freshly distilled, degassed Ac$_2$O under $N_2$. The suspension was placed in a 150 °C preheated oil bath and stirred for 3.5 h. The solution was cooled to 24 °C and filtered. A mixture of MeOH/H$_2$O (2/1) was added to the filtrate and the brown solid was filtered. The brown solid was subjected to column chromatography (SiO$_2$; 1:4 DCM/hexanes) yielding a light yellow solid. The crude material (61 mg) was added to Ac$_2$O (1.5 mL) and toluene (0.5 mL). The reaction mixture was cooled to 0 °C and HBF$_4$ (0.1 mL, 48 wt.% in H$_2$O) was added. The reaction mixture turned orange and was stirred at 0 °C for 10 min. The reaction mixture was added to anhydrous Et$_2$O and filtered. The solvent was removed from the filtrate and the resulting solid was sonicated in cold MeOH then filtered and washed with cold MeOH yielding a light yellow solid. The solid was further purified via four recrystallizations from EtOH/CHCl$_3$ yielding **1** (5.3 mg, 0.007 mmol, <1 %) as a light yellow powder. $^1$H NMR (400 MHz, CD$_2$Cl$_2$, 22 °C) δ = 8.73

(d, *J* = 8.9, 2H), 7.98 (d, *J* = 8.1, 1H), 7.71 (dd, *J* = 7.8, 1.5, 2H), 7.64–7.59 (m, 3H), 7.55 (d, *J* = 8.7, 2H), 7.49 (dd, *J* = 8.1, 2.2, 1H), 7.44 (ddd, *J* = 7.9, 7.1, 1.1, 2H), 7.41–7.20 (m, 11H), 7.17 (ddd, *J* = 8.7, 7.0, 1.6, 2H), 6.84 (d, *J* = 9.1, 2H) ppm; $^{13}$C {$^{1}$H} NMR (151 MHz, CD$_2$Cl$_2$, 22 °C) δ = 145.4, 145.3, 141.4, 137.8, 135.9, 135.6, 135.0, 134.5, 133.1, 133.0, 132.9, 132.0, 131.5, 131.1, 130.3, 129.9, 128.9, 128.8, 128.6, 128.6, 128.5, 128.3, 127.9, 127.7, 127.0, 125.2, 124.9, 124.2, 122.8 ppm; HRMS (EI-TOF) m/z: [C$_{48}$H$_{28}$Br$_2$]$^+$ calcd. [C$_{48}$H$_{28}$Br$_2$] 764.0537; found 764.0539.

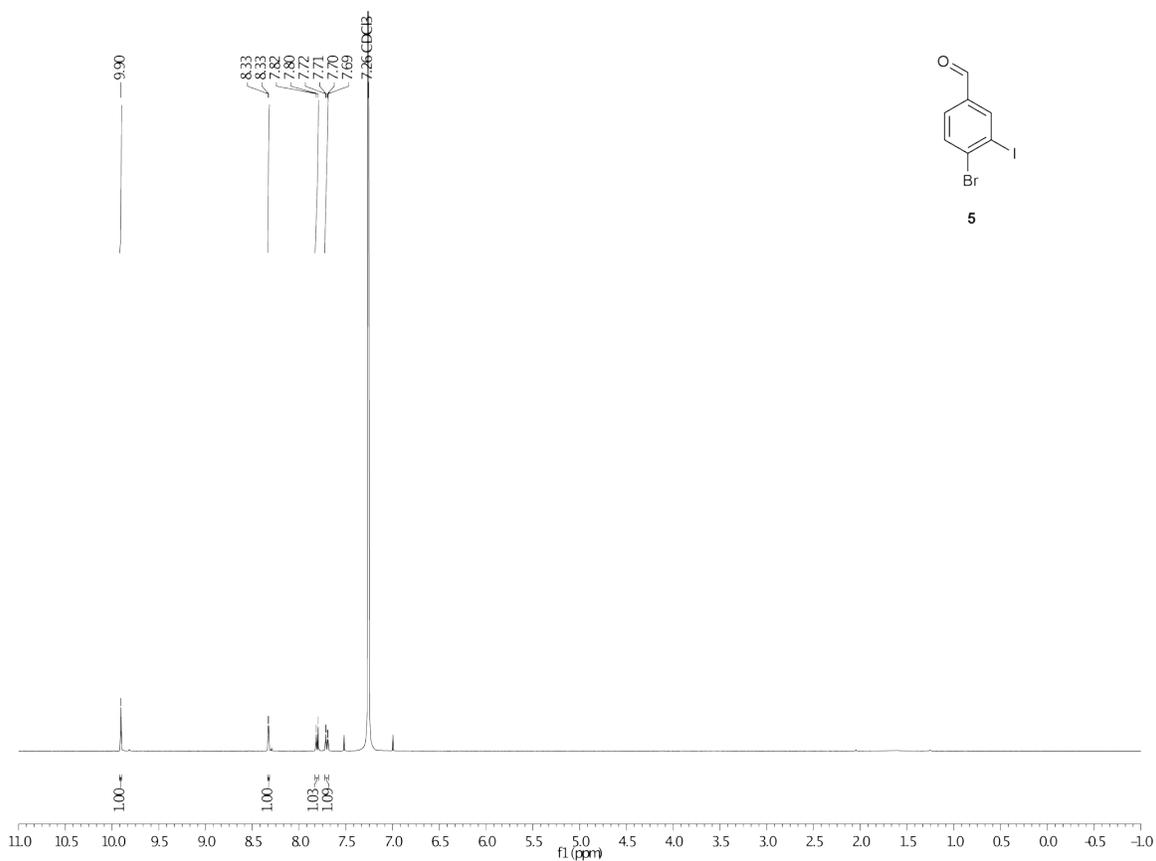

**Supplementary Figure S5** | $^1$H NMR (400 MHz, 22 °C, CDCl$_3$) of **5**.

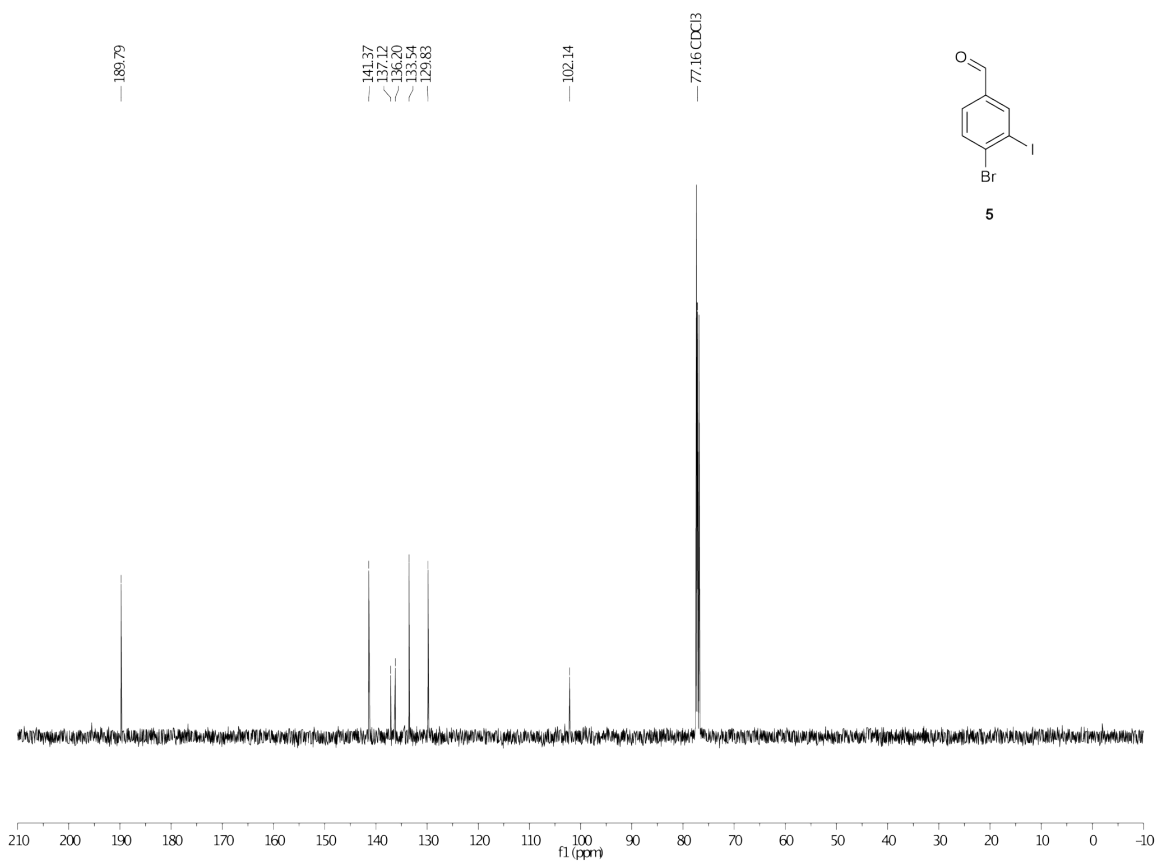

**Supplementary Figure S6** | $^{13}$C NMR (101 MHz, 22 °C, CDCl$_3$) of **5**.

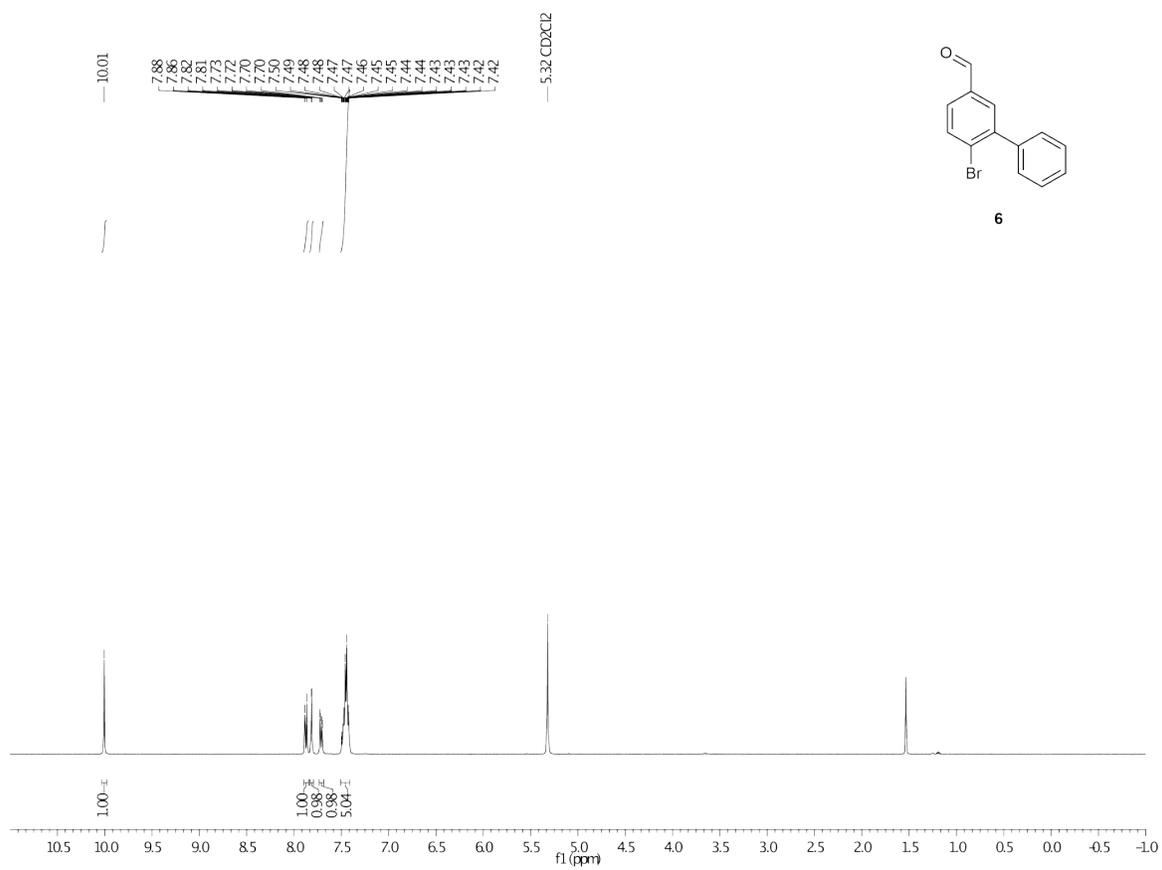

**Supplementary Figure S7** | $^1$H NMR (400 MHz, 22 °C, CD$_2$Cl$_2$) of **6**.

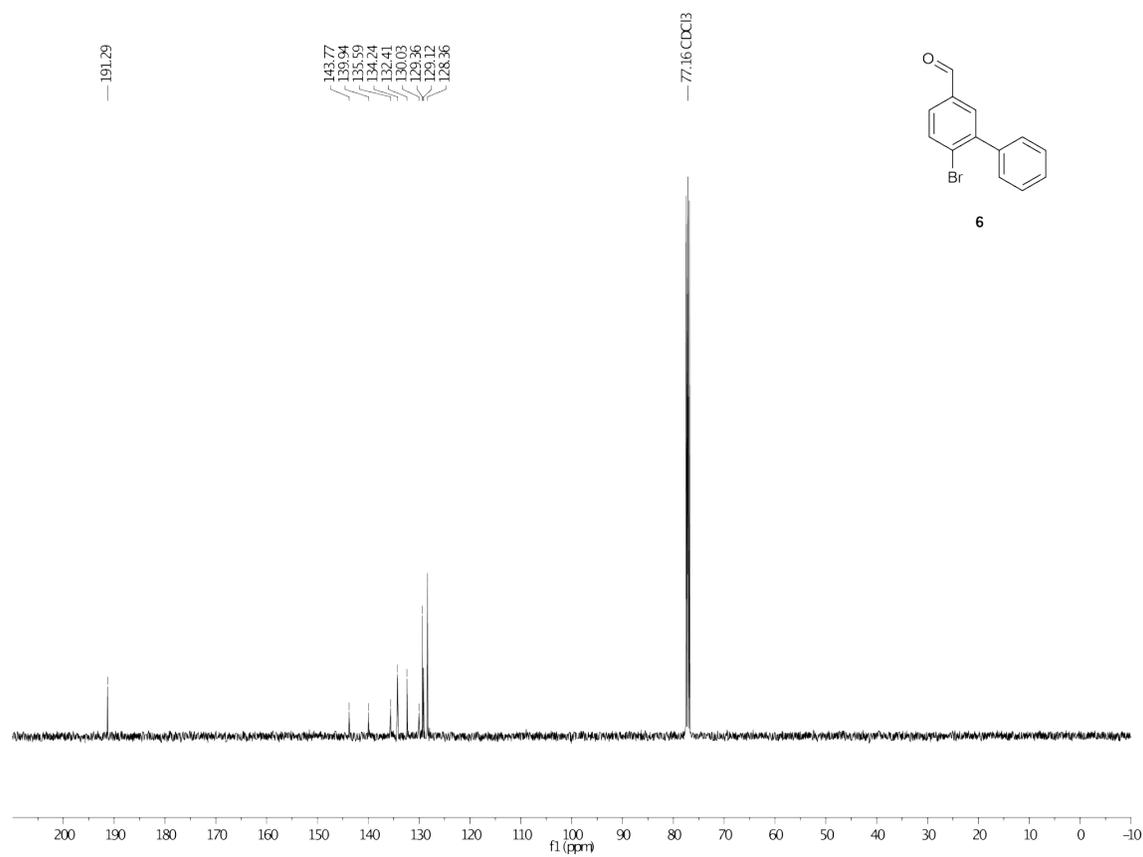

**Supplementary Figure S8** | $^{13}$C NMR (151 MHz, 22 °C, CDCl$_3$) of **6**.

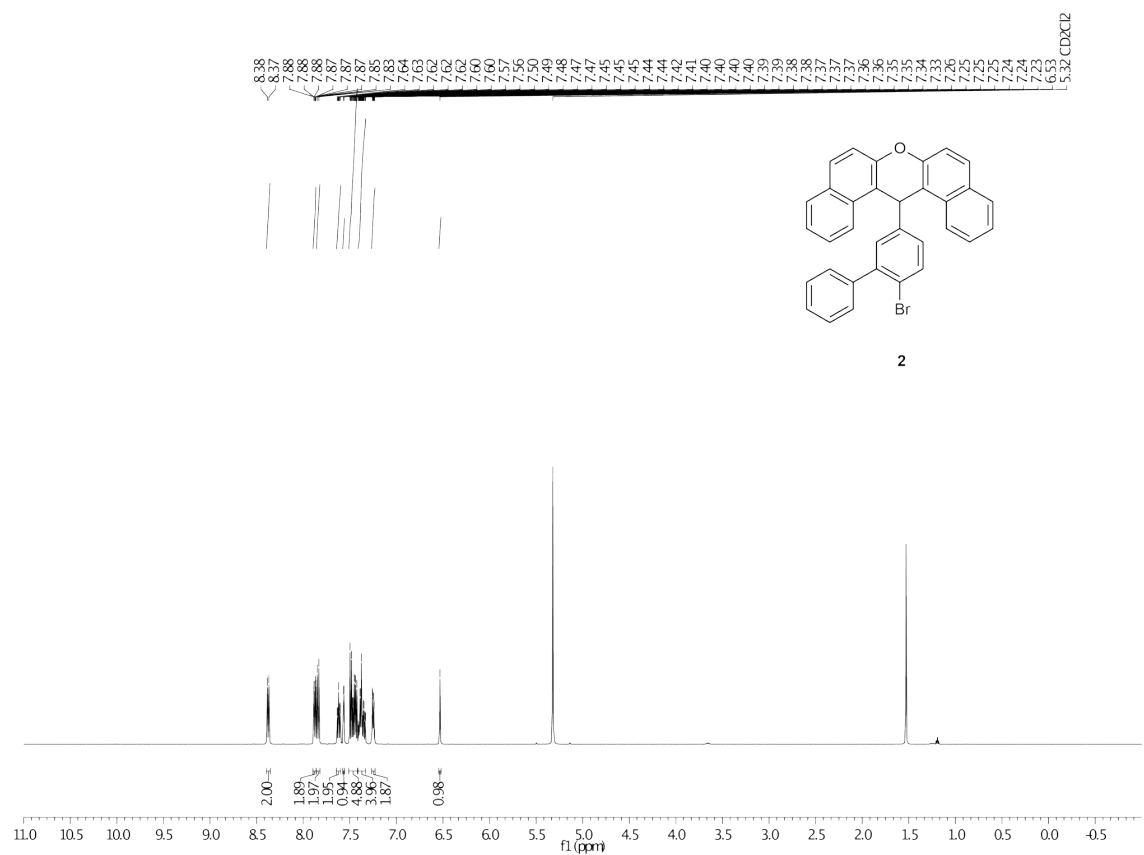

**Supplementary Figure S9** | $^1$H NMR (500 MHz, 22 °C, CD$_2$Cl$_2$) of **2**.

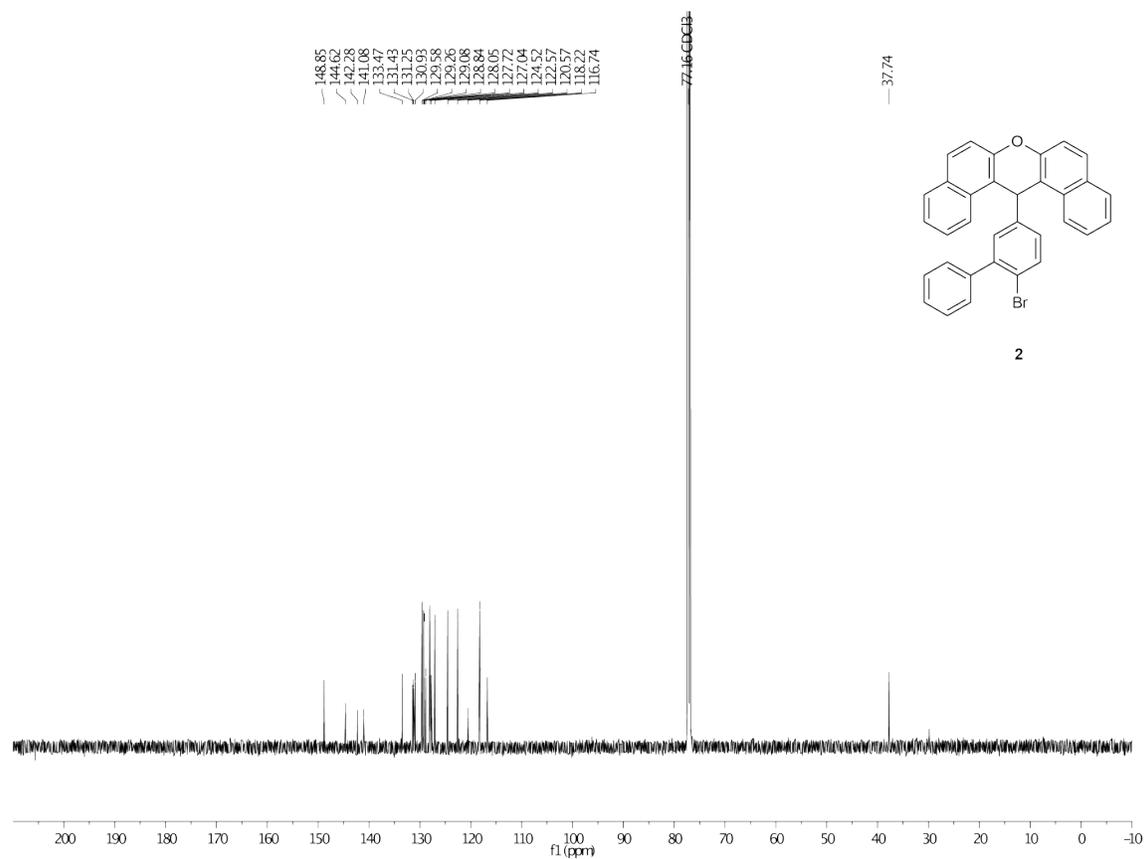

**Supplementary Figure S10 |** $^{13}$C NMR (151 MHz, 22 °C, CDCl$_3$) of **2**.

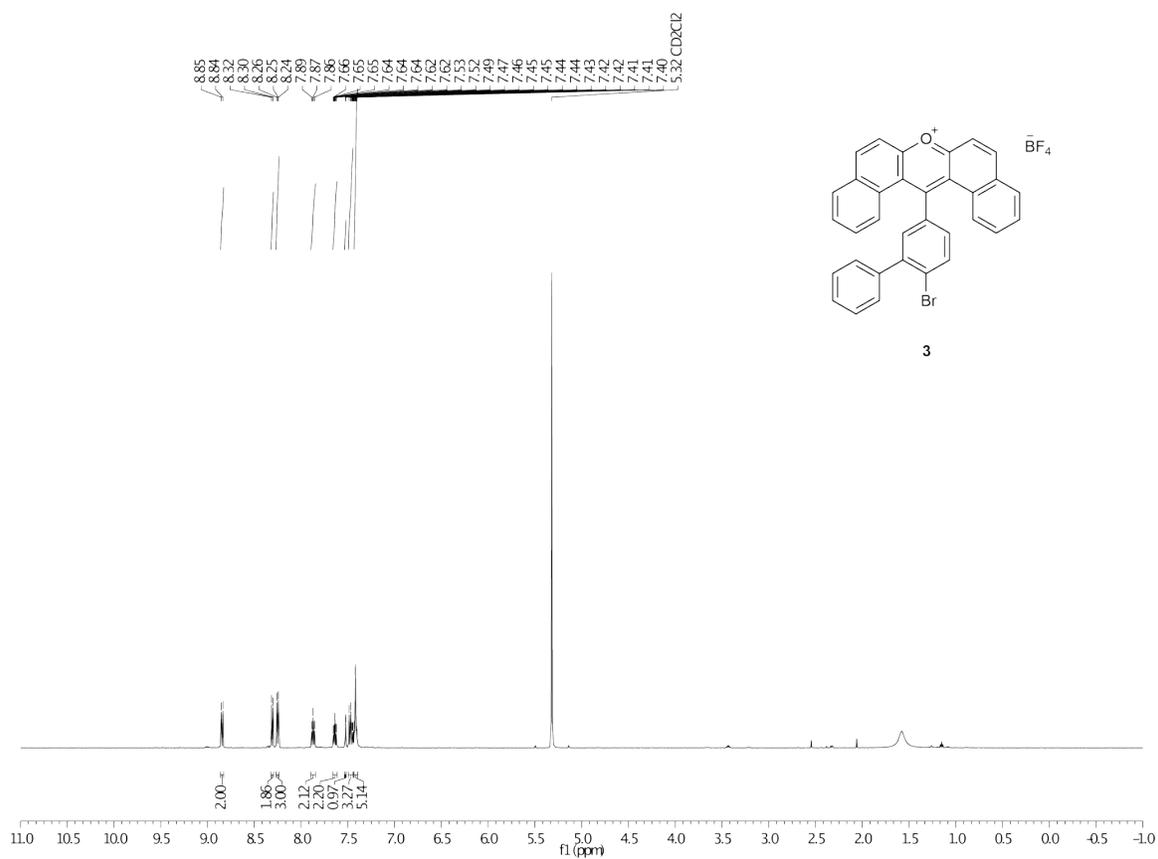

**Supplementary Figure S11** | $^1$H NMR (500 MHz, 22 °C, CD$_2$Cl$_2$) of **3**.

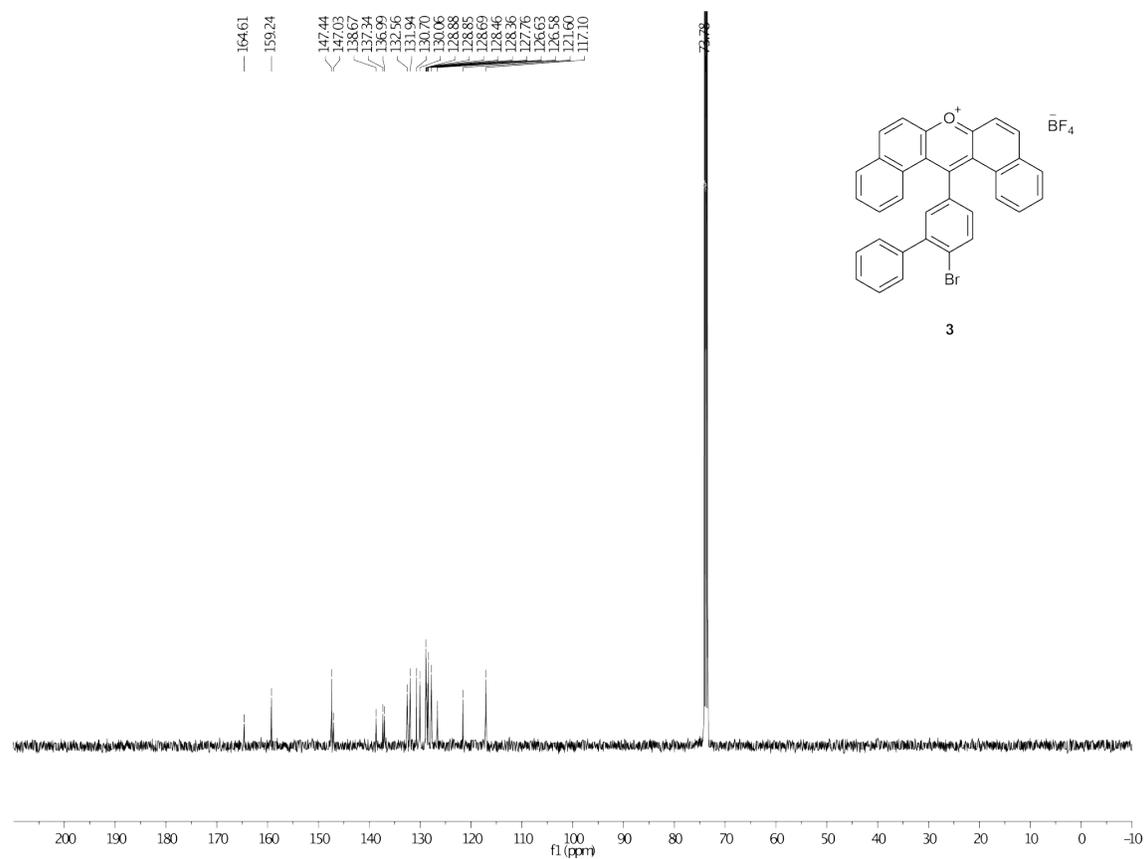

**Supplementary Figure S12** | $^{13}$C NMR (101 MHz, 22 °C, C$_2$D$_2$Cl$_4$) of **3**.

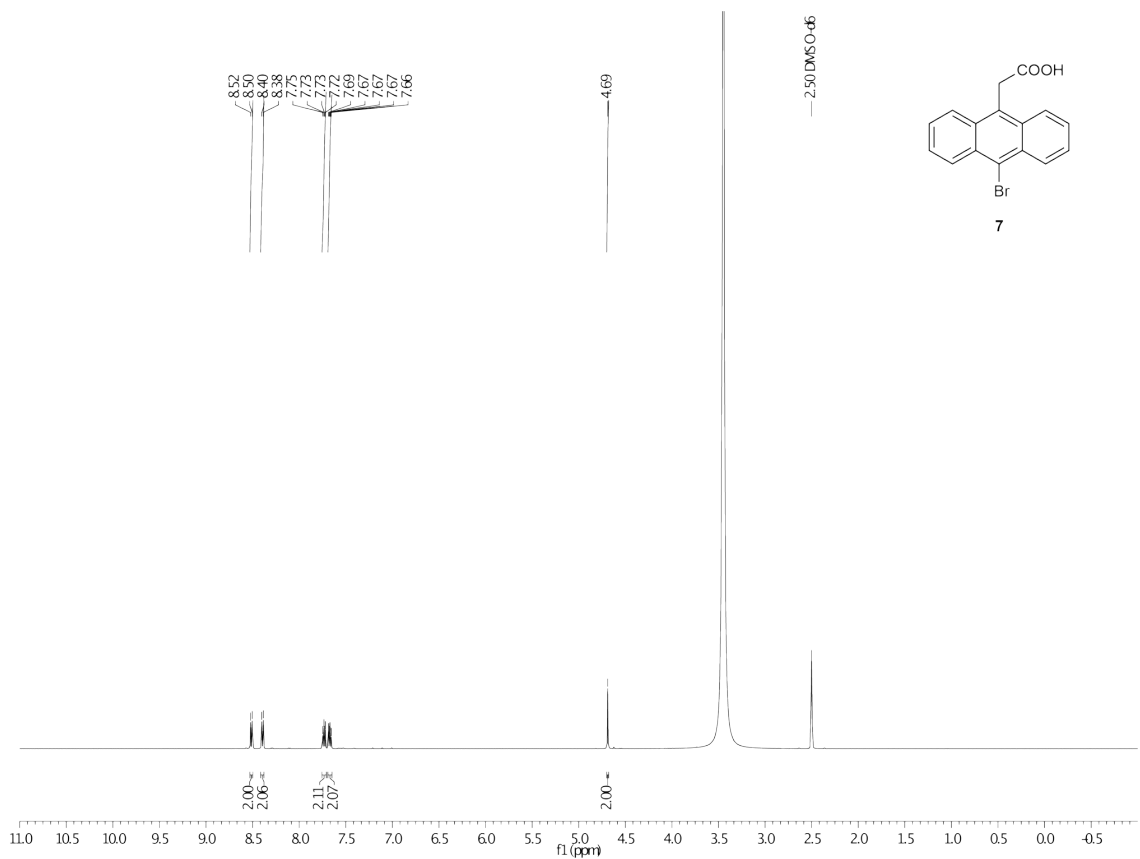

**Supplementary Figure S13** | $^1$H NMR (500 MHz, 22 °C, DMSO-$d_6$) of **7**.

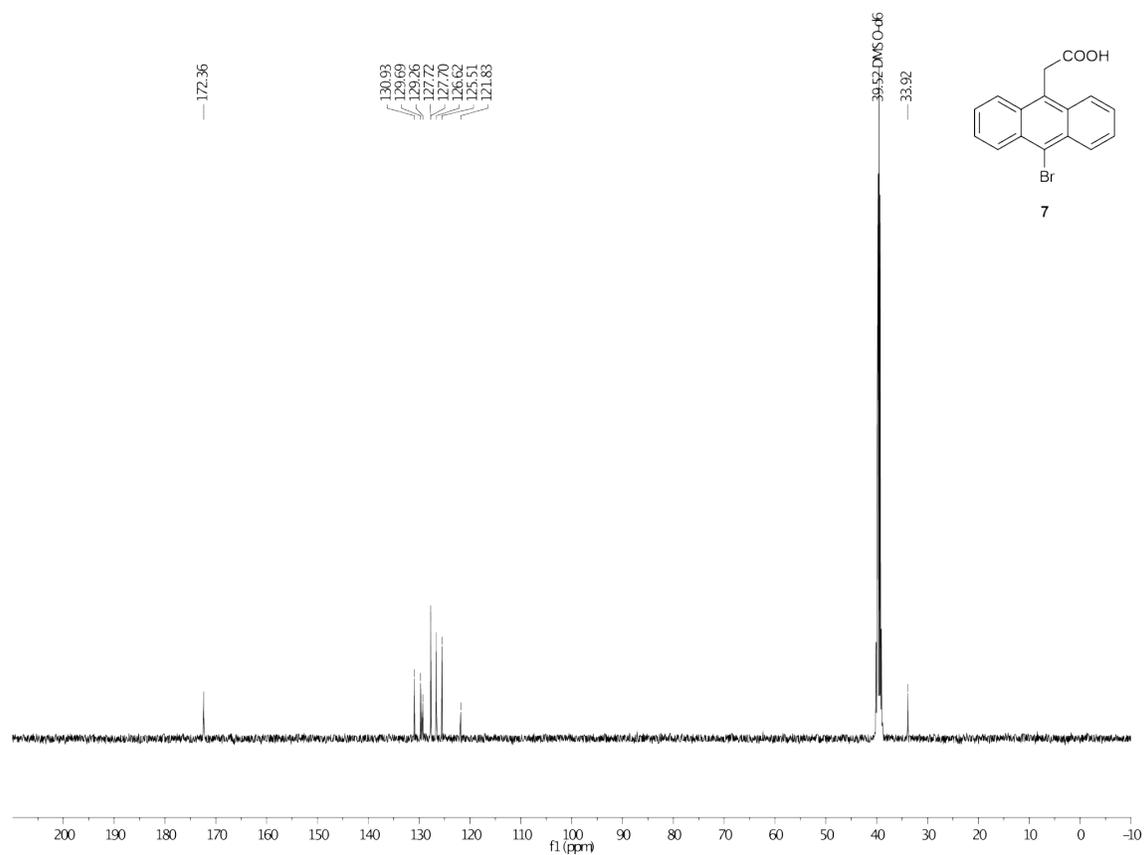

**Supplementary Figure S14** | $^{13}$C NMR (101 MHz, 22 °C, DMSO-$d_6$) of **7**.

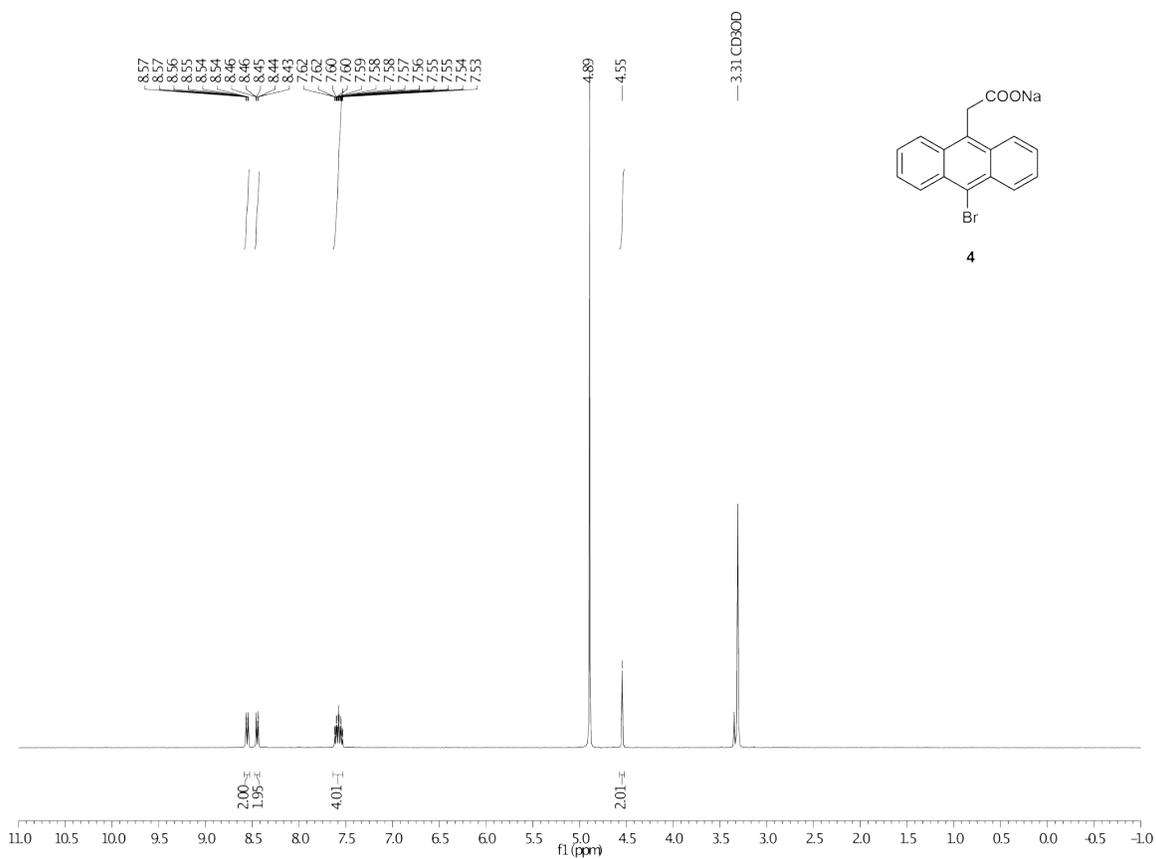

**Supplementary Figure S15** | $^1$H NMR (400 MHz, 22 °C, CD$_3$OD) of **4**.

**Supplementary Figure S16** | $^1$H NMR (400 MHz, 22 °C, CD$_2$Cl$_2$) of **1**.

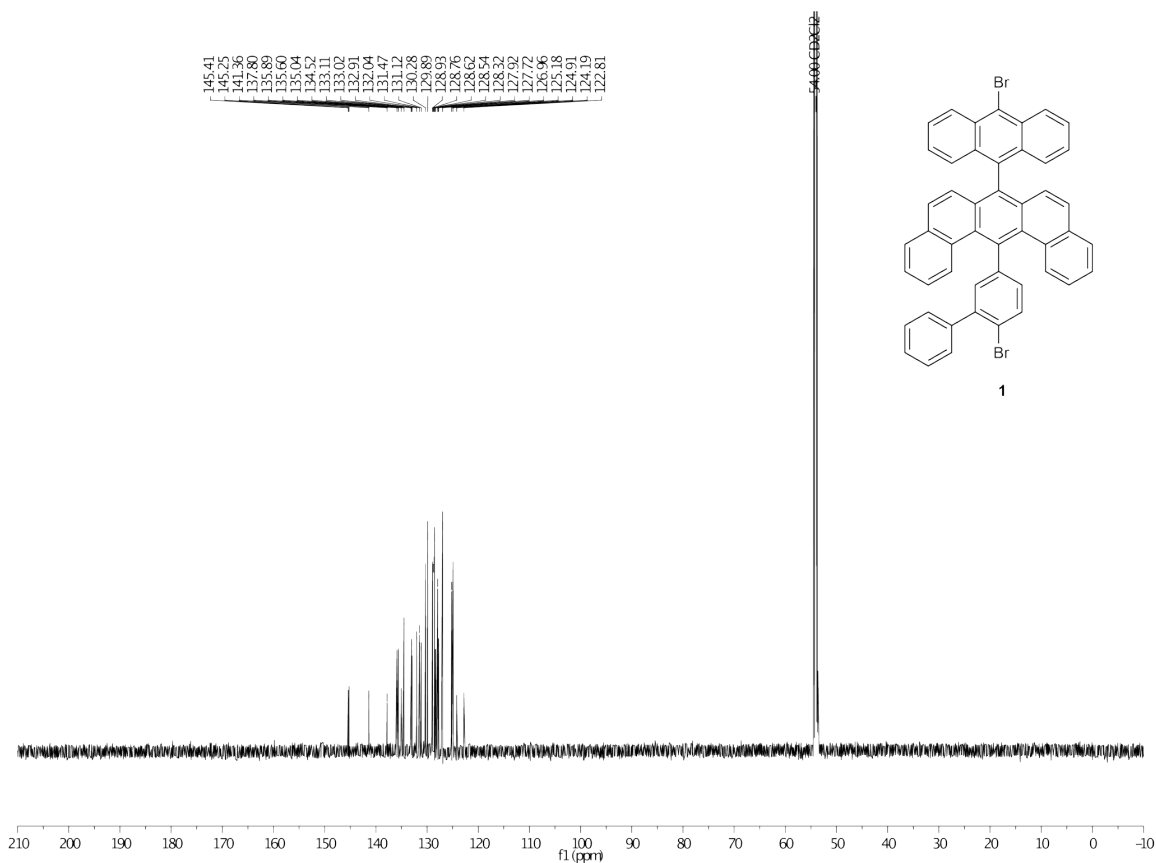

**Supplementary Figure S17** | $^{13}$C NMR (151 MHz, 22 °C, CD$_2$Cl$_2$) of **1**.